# The Joint Efficient Dark-energy Investigation (JEDI): Measuring the cosmic expansion history from type Ia supernovae

M. M. Phillips[*a], Peter Garnavich[b], Yun Wang[c], David Branch[c], Edward Baron[c], Arlin Crotts[d], J. Craig Wheeler[e], Edward Cheng[f], and Mario Hamuy[g] for the JEDI Team

[a]Carnegie Observatories, Las Campanas Observatory, Casilla 601, La Serena, Chile;
[b]Dept. of Physics, Univ. of Notre Dame, Notre Dame, IN 46556, USA;
[c]Dept. of Physics & Astronomy, Univ. of Oklahoma, 440 W. Brooks St., Norman, OK 73019, USA;
[d]Columbia University, Astrophysics Laboratory, 550 W. 120th St., New York, NY 10027, USA;
[e]Dept. of Astronomy, Univ. of Texas, Austin, TX 78712, USA;
[f]Conceptual Analytics, LLC, 8209 Woburn Abbey Rd., Glenn Dale, MD 20769, USA;
[g]Depto. de Astronomía, Universidad de Chile, Casilla 36-D, Santiago, Chile

## ABSTRACT

JEDI (Joint Efficient Dark-energy Investigation) is a candidate implementation of the NASA-DOE Joint Dark Energy Mission (JDEM). JEDI will probe dark energy in three independent methods: (1) type Ia supernovae, (2) baryon acoustic oscillations, and (3) weak gravitational lensing. In an accompanying paper, an overall summary of the JEDI mission is given. In this paper, we present further details of the supernova component of JEDI. To derive model-independent constraints on dark energy, it is important to precisely measure the cosmic expansion history, $H(z)$, in continuous redshift bins from $z \sim 0$-$2$ (the redshift range in which dark energy is important). SNe Ia at $z > 1$ are not readily accessible from the ground because the bulk of their light has shifted into the near-infrared where the sky background is overwhelming; hence a space mission is required to probe dark energy using SNe. Because of its unique near-infrared wavelength coverage (0.8-4.2 μm), JEDI has the advantage of observing SNe Ia in the rest frame $J$ band for the entire redshift range of $0 < z < 2$, where they are less affected by dust, and appear to be nearly perfect standard candles. During the first year of JEDI operations, spectra and light curves will be obtained for ~4,000 SNe Ia at $z < 2$. The resulting constraints on dark energy are discussed, with special emphasis on the improved precision afforded by the rest frame near-infrared data.

## 1. INTRODUCTION

The expansion of the Universe is accelerating[1,2,3,4,5,6], driven by an unknown mechanism that is referred to as "Dark Energy". The equation of state of the Dark Energy can be written as $w(z) = w_0 + w'z$. There are two classes of possibilities for the nature of the Dark Energy. It could be an unknown energy component with density $\rho_X(z)$ in the Universe ($\rho'_X(z)/\rho_X(z) = 3(1+w(z))/(1+z)$), a static "cosmological constant" with density constant in time ($\rho_X(z) = \rho_X(0)$; $w_0=-1$ and $w'=0$), or a dynamical quantity with density $\rho_X(z)$ that varies with cosmic time. Alternatively, Dark Energy could be the consequence of deviations from General Relativity (hence not energy, but modified gravity). Solving the mystery of Dark Energy is arguably the most important problem in cosmology today.

In response to this scientific need, the Joint Dark Energy Mission (JDEM) is being sponsored by NASA and DOE to investigate the nature of Dark Energy in our Universe. JEDI (Joint Efficient Dark-energy Investigation) is a candidate implementation of the JDEM. The nature of dark energy can be probed with different cosmological observations: the cosmic expansion history $H(z)$ (the expansion rate of the Universe as a function of $z$); luminosity- and angular-diameter-distances as functions of $z$, $d_L(z)$ and $d_A(z)$, respectively; and the linear growth factor of cosmic large scale structure, $G(z)$ (the growth rate of matter density perturbations in the Universe relative to that of a flat universe without Dark Energy). The JEDI mission concept provides measurements over a large redshift range, $0 < z < 2$, through three powerful and complementary observational surveys. Through the combination of a Deep and Wide campaign, we survey a) Type Ia supernovae as standard candles ($d_L(z)$); b) cosmic shear as a tracer of weak lensing from intervening matter which gives $d_A(z_1)/d_A(z_2)$ ratios and the cosmic structure growth factor $G(z)$; and c) baryon acoustic oscillations in the galaxy distribution as a standard ruler in both the transverse and radial directions ($d_A(z)$ and $H(z)$), and a probe of $G(z)$. Though all of these measures are formally related for a given cosmological model, they all suffer from different systematic uncertainties. The JEDI mission concept is designed to have the complementarity and redundancy that is essential to achieve not only extremely small stochastic errors, but to identify and account for the systematic errors to which each method is susceptible. This strategy minimizes the scientific risk to such a mission. To fully exploit each of these methods to the levels of the known astrophysical limits, the JEDI mission will need to achieve extremely deep and uniform imaging and spectroscopy at infrared wavelengths not accessible from the ground. These requirements, combined with the need for a small and extremely stable point-spread function (PSF) drives the need for a space-based platform.

---

[*] mmp@lco.cl

In an accompanying paper, an overall summary of the JEDI mission is given. In this paper, we present further details of the supernova component of JEDI. Dark Energy was first discovered through observations of Type Ia supernovae (SNe Ia)[1,2], and this method continues to be one of the most powerful and attractive for direct measurement of the cosmic expansion history as a function of redshift.

## 2. TYPE IA SUPERNOVAE AS COSMOLOGICAL STANDARD CANDLES

Since the pioneering observations of Baade in the 1930's and 1940's, the light curves of SNe Ia have been known to be remarkably homogeneous[7]. However, the utility of SNe Ia as cosmological standard candles was not fully demonstrated for another 50 years when several studies[8,9,10,11,12] yielded Hubble diagrams with dispersions within the observational errors (0.3-0.5 mag), suggesting that SNe Ia might actually have identical light curves and peak luminosities[13]. During approximately the same period of time, a different point of view was advocated by Pskovskii[14,15,16] who argued that SNe Ia display a continuous range of decline rates which correlate with several different parameters, including absolute magnitude. Pskovskii found that the absolute magnitudes of SN Ia were inversely correlated with the decline rate - i.e., events that decline more rapidly from maximum are less luminous than those which decline more slowly. While there was some support for this idea[17], in general his conclusions were not embraced by others in the field.

This issue was not settled until the implementation of CCD photometry techniques beginning in the late 1980s. The first ever CCD light curves published for a SN Ia were for SN 1986G in NGC 5128[18]. These demonstrated unequivocally that this supernova declined significantly faster from maximum than the well-observed SN 1981B for which precise photoelectric photometry had been obtained. A few years later, another well-observed, fast-declining event, SN 1991bg, appeared in the Virgo cluster member M84 and was found to be significantly under-luminous[19,20], lending credence to a Pskovskii-type correlation between luminosity and decline rate. These findings motivated a re-examination of the issue by Phillips[21] who employed a sample of nine well-observed SNe Ia with host galaxy distances estimated via the surface brightness fluctuations or Tully-Fisher methods. As a convenient measure of the decline rate, Phillips introduced the parameter $\Delta m_{15}(B)$, defined as the amount in magnitudes that the $B$ light curve declines during the first 15 days following maximum. This study demonstrated the existence of a significant intrinsic dispersion in the absolute magnitudes at maximum light of SNe Ia, amounting to ~0.8 mag in $B$, ~0.6 mag in $V$, and ~0.5 mag in $I$. Moreover, the absolute magnitudes were correlated with the initial rate of decline of the $B$ light curve, with the slope of the correlation being steepest in $B$ and becoming progressively flatter in the $V$ and $I$ bands.

While the discovery of a significant dispersion in the absolute magnitudes of SNe Ia complicated the use of these objects as cosmological standard candles, the smoothness of the correlation with decline rate offered the possibility of applying a luminosity correction in analogy with the period-luminosity relation for Cepheid variables. During the rest of the 1990s, considerable effort was expended on determining the exact dependence of the absolute magnitudes on the decline rate. In 1996, the Calán/Tololo project produced a Hubble diagram with the lowest dispersion ever achieved of only 0.14 mag, which firmly established that distance determinations to 7% or better were possible using SNe Ia[22].

Figure 1 shows a modern version of the absolute magnitude vs. decline rate relations in $BVIJHK$ for SNe Ia. These data confirm that the slope of the correlation between absolute magnitude and decline rate is greatest in the $B$ band, and becomes progressively flatter at redder wavelengths. In the near-infrared, the slope of the relations can actually be neglected for SN Ia with $\Delta m_{15}(B) < 1.4$, confirming a long-held suspicion[21,23,24] that these objects are essentially perfect standard candles in the $JHK$ bands. The principal advantages of the near-infrared are two-fold: reddening corrections are typically vanishingly small, and luminosity corrections can essentially be ignored except for the very fastest-declining events. The combination of optical and near-infrared photometry also yields much more reliable dust extinction corrections than can be obtained from optical data alone[25]. A recent study of 16 SNe Ia observed in the near-infrared yielded Hubble diagrams in *JHK* with a dispersion of ~0.15 mag (~7% in distance) without making any luminosity correction whatsoever[26].

## 3. THE JEDI SUPERNOVA SURVEY

To derive model-independent constraints on dark energy using SNe, it is important that we precisely measure the cosmic expansion history, $H(z)$, in continuous redshift bins from $z \sim 0\text{-}2$ (the redshift range in which dark energy is important). SNe Ia at $z > 1$ are not accessible from the ground, because the bulk of their light has shifted into the near-infrared where the sky background is overwhelming; hence a space mission is required to probe dark energy using SNe.

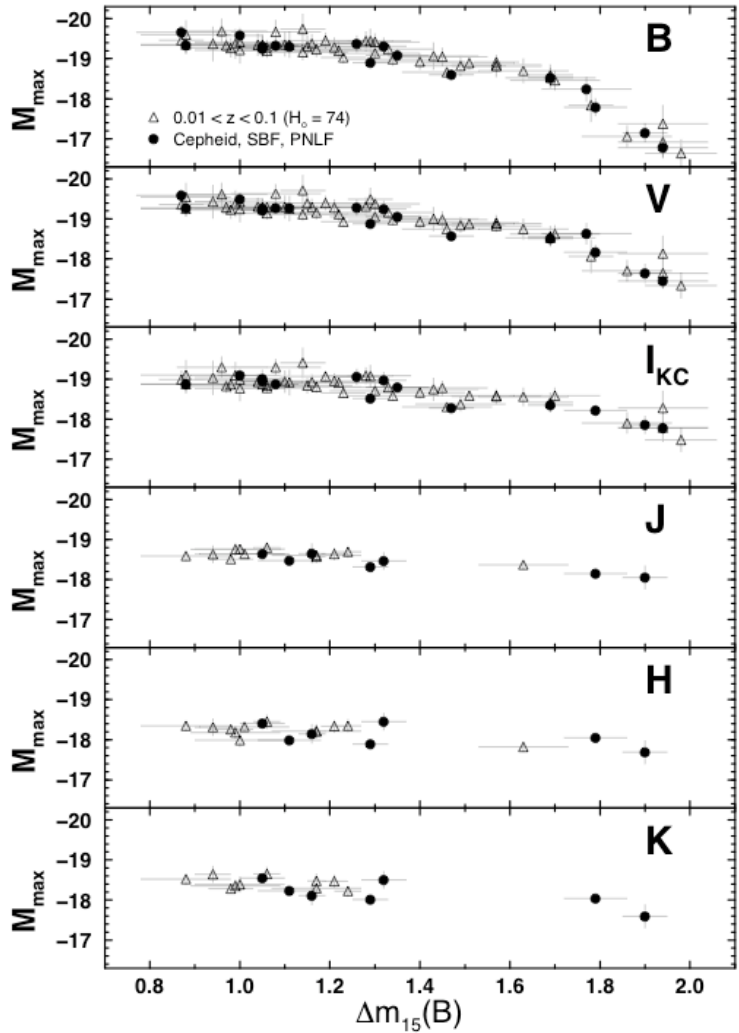

Fig. 1. Absolute magnitude vs. decline rate relation in *BVIJHK* for two samples of SNe Ia. The first consists of 17 well-observed nearby SNe Ia which occurred in host galaxies for which distances have been derived via Cepheids, surface brightness fluctuations, or the planetary nebula luminosity function method. The second sample consists of 50 SNe Ia in the Hubble flow ($z > 0.01$). Distances for these objects were derived from the host galaxy recession velocity assuming $H_0 = 74$ kms$^{-1}$ Mpc$^{-1}$ which gives the best agreement between the two samples. Note that all of the data have been corrected for both Galactic and host galaxy reddening.

The JEDI SNe Ia survey will be executed during the first year of the three-year mission as part of the JEDI Deep Campaign. The baseline mission is to measure ~4,000 SNe Ia ($0 < z < 2$) with light curves sampled every 7 days (3-4 days in the supernova rest frame) and spectra with S/N ≥ 10. The JEDI Deep Campaign will accomplish this by continuously monitoring a 4 deg$^2$ field at the north ecliptic pole with a cadence of 7 days. The focal plane instrumentation is built for five-band photometric imaging from 0.8-4.2 μm, with simultaneous multi-object spectroscopy covering the wavelength range 1-2 μm in an adjoining field. The imaging part of the survey will reach a magnitude limit of $H_{AB}$ = 26.5 at S/N = 10 at each visit. This is sensitive enough for observing SNe Ia at $z = 3$ (if they exist at such high redshift). Our preliminary simulations indicate that while we will have very high quality SN Ia spectra at low and intermediate redshifts, we will have good quality SN Ia spectra (S/N ≥ 10) at $z = 1.6$ and beyond, which will be improved by co-adding the spectra from successive visits. Co-adding spectra of a SN Ia obtained at multiple epochs near maximum light preserves not only the SN Ia spectroscopic signature, but also the specific signature of the SN Ia subtype[27].

A precision measurement of $H(z)$ from SNe requires systematic uncertainties to be controlled to better than the 2% level. Calibration is a critical issue for the entire JEDI mission, but this is especially true for the SNe photometry. We anticipate requiring 1-2% absolute calibration to tie in with low-redshift, ground-based data, and an intra-band calibration accuracy to 0.5-1.0%. Systematic errors can also be lurking in the SNe data themselves. If the observed scatter of ~10% in SNe Ia-determined distances in the local Universe were caused by a random process such as the viewing angle of slightly asymmetric explosions, then reaching the desired error floor of 2% would be a trivial average of all SNe Ia at a given redshift. However, there are other effects which may be a source of dispersion but vary with cosmic time and may result in systematic error. These include:

**Dust Extinction**
Currently extinction by host galaxy dust is estimated by knowing the intrinsic color of SNe Ia and applying a standard reddening law. There is some controversy whether the standard Galactic reddening law applies and a variation in the host dust properties can contribute to the observed dispersion. Observing over a wide range of wavelengths that include the near-infrared, where dust extinction is less important, allows an independent estimate of the extinction law. Moreover, we will observe the rest-frame $J$ band light curves for all the JEDI SNe Ia by having a mildly cryogenic telescope that reaches 4.2 μm. At this wavelength, dust extinction can essentially be ignored since $A_J / A_V = 0.28$. Hence, for the JEDI SNe Ia, the overall effect of extinction can be reduced to less than 1%.

**K-Corrections**
Photometry of the SNe Ia at different redshifts in different color bands needs to be mapped onto a consistent rest-frame band. We will use a set of rest-frame template spectra to build a rest-frame Spectral Energy Distribution (SED) for the entire SN Ia wavelength range. Because of the extraordinary efficiency of the JEDI spectrographs, we will have at least several high signal-to-noise ratio, moderate resolution spectra per SN Ia at different epochs. This will enable us to calibrate and improve our template SED, and reduce the uncertainty in K-corrections to around 0.01 mag. Furthermore, the JEDI bands are spaced in wavelength so that they transform into each other at certain redshifts, so that uncertainties in K-correction identically disappear. Finally, the redshift dependent systematic bias due to K-correction uncertainties can be reduced to a negligible level by optimizing the analysis technique; Wang & Tegmark[28] have developed a method that effectively reduces the global systematic bias (over the entire redshift range) to a local bias (in each redshift bin) with a much smaller amplitude.

**Weak Lensing**
The bending of the light from SNe Ia by intervening matter will modify the observed brightness of SNe Ia. It has been demonstrated that the effect of weak lensing of SNe Ia can be reduced to an negligible level given sufficiently good statistics[28,29,30]. For JEDI, this will be smaller than 0.2%.

**Selection Bias**
Krisciunas et al.[31] has shown the importance of searching at a redshift limit, not a magnitude limit. To avoid Malmquist-like biases, it is necessary to search a magnitude deeper than the typical peak supernova brightness to sample the fast-decaying events. JEDI exposure times are designed to detect fainter-than-typical events at $z = 2$.

**Supernova Peak Luminosity Evolution**

Models suggest that progenitor metallicity or mass may influence the peak luminosity of SNe Ia at a fixed light curve decline rate, although, the models do not agree on the amplitude or even the sign of these effects. Since both progenitor population metallicity and age vary over cosmic time this is an important and uncertain source of systematic error. Progenitor metallicity may be correlated with ultraviolet flux before maximum, but no observations have demonstrated this effect. There is a strong correlation between light curve decline rate and host morphology (and host star-formation rate)[32], but Gallagher et al.[33] have shown that there is no strong residual correlation between host metallicity or star-formation history after correction for light curve shape and dust extinction. With the current uncertain state of theory and observation, our plan is to obtain a sufficient range and details in our observations to allow a study of environmental influences at all redshifts. With the 4,000 SNe Ia with well sampled light curves and good quality spectra from JEDI, we will be able to subtype the SNe Ia, and search for most possible evolutionary effects.

## 4. EXPECTED CONSTRAINTS ON DARK ENERGY

For the baseline sample of 4,000 SNe Ia ($0 < z < 2$) observed during the first year of JEDI, the expected $1\sigma$ errors on $w_0$ and $w'$ are 0.045 and 0.152, respectively. However, the challenge to the Dark Energy parameter measurements will not be with precision (from statistics), but in determining the accuracy, which requires understanding and adequate modeling of systematic uncertainties. Multiple observational methods with different systematic susceptibilities resolve this problem, and drive the entire JEDI mission rationale. Figure 2 illustrates the advantage of the JEDI mission approach. Measurement-error ellipses from each of three techniques - supernovae (SNe), baryon oscillations (BAO), and weak lensing (WL) - are tilted in different directions; this is because the three JEDI methods probe Dark Energy in different ways, resulting in the combined error ellipse shown. If unaccounted-for systematic effects dominate, the error ellipses may be offset from each other, and the goodness-of-fit of the results would be poor. With the JEDI mission redundancy, these systematic effects can be identified, analyzed, and removed.

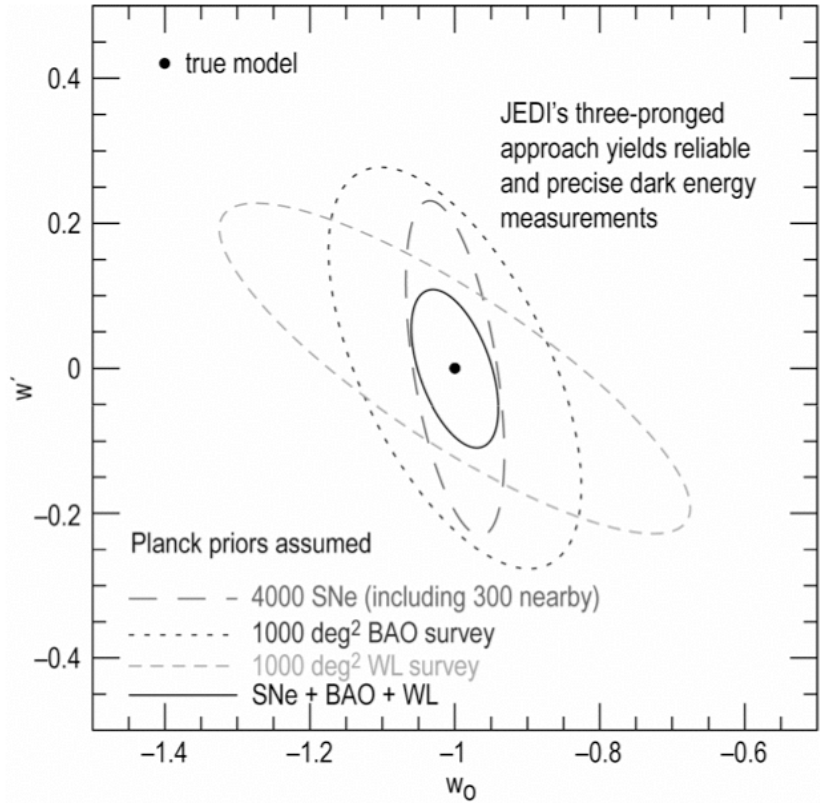

Fig. 2. Dark Energy measurements expected from the JEDI baseline mission, showing $1\sigma$ (68.3%) confidence level contours. Each error ellipse becomes offset from the true model ($\Omega=0.3$, $w=-1$, $w'=0$) when systematics dominate.

The lower panel of Figure 3 shows the expected precision with which the JEDI mission will measure $H(z)$ for $0 < z < 2$ using both SNe and BAO, thus enabling model-independent constraints on the time dependence of Dark Energy[28]. For comparison, a plot of $H(z)$ recently derived by Wang and Mukherjee[34] from the best current data[5,6] is shown in the upper panel of Figure 3. The JEDI WL survey will provide independent $H(z)$ measurements[35], and serve as a cross-check of the SNe and BAO results. Note that these simulations assume the baseline sample of 4,000 SNe Ia which is conservatively defined to match modest current data handling capabilities. The expected increase in data handling capabilities prior to the JDEM launch date will allow JEDI to obtain ~$10^4$ SNe and execute $10^4$ deg$^2$ WL and BAO surveys.

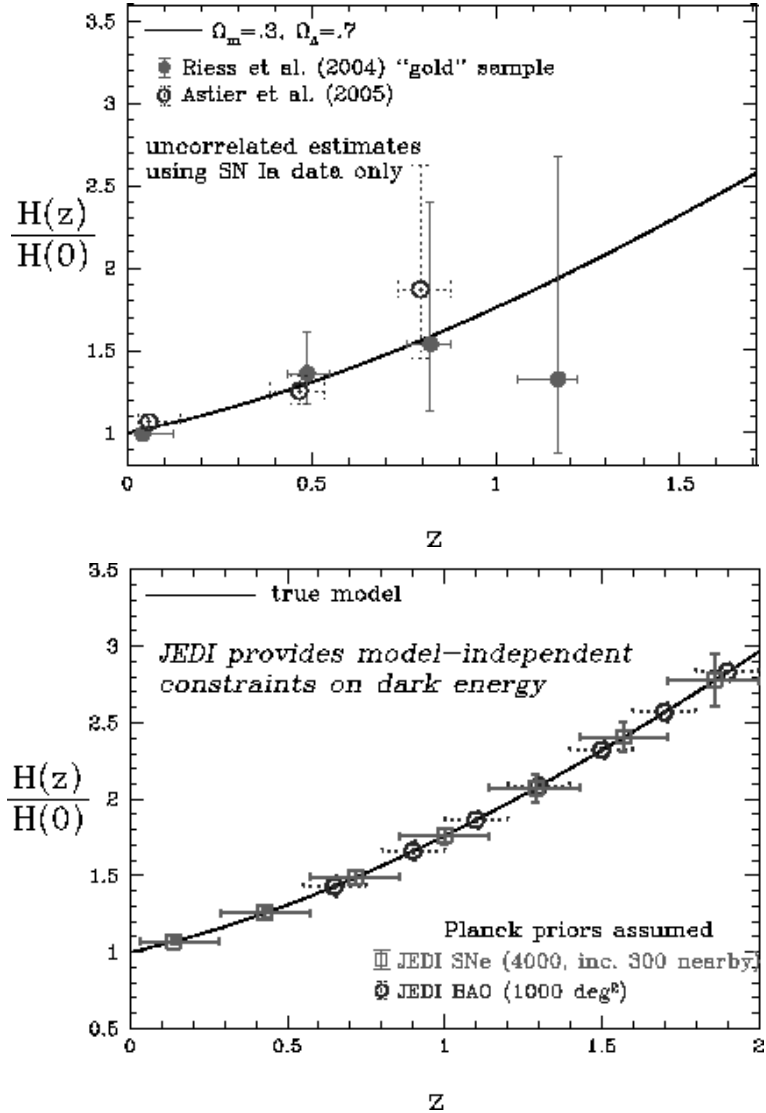

Fig. 3. Upper panel: The uncorrelated cosmic expansion history $H(z)/H(z=0)$ estimated using the best current SNe Ia data. Lower panel: Expected $H(z)/H(z=0)$ from the JEDI baseline observations of SNe and BAO, with 1σ errors. Note that the errors go opposite ways in the two methods.

The simulations illustrated in Figures 2 and 3 include an assumed sample of 300 nearby ($z < 0.1$) SNe Ia observed in ongoing ground-based programs. Chief among these are the Carnegie Supernova Project[36] (CSP), which will obtain $u'g'r'i'YJHK$ light curves of ~100 SNe Ia at $z < 0.07$, and the Supernova Factory[37], which will discover and obtain frequent spectrophotometry covering 3200-10000 Å of ~300 SNe Ia at $0.03 < z < 0.08$.

## 5. CONCLUSIONS

The proposed JEDI mission will use SNe Ia to precisely measure the cosmic expansion history, $H(z)$, in continuous redshift bins from $z \sim 0\text{-}2$. JEDI achieves extraordinary efficiency by providing for 5-band imaging to be carried out while simultaneously obtaining slit spectra of previously-identified SNe Ia. In addition, JEDI uniquely exploits near-infrared wavelength coverage (0.8-4.2 μm) to measure SNe Ia in the rest frame $J$ band over this entire redshift range, where they are less affected by dust, and appear to be nearly perfect standard candles. By combining the SNe data with baryon oscillation and weak lensing observations obtained in a wide field (1,000 deg$^2$) survey to take place during years two and three of the mission, JEDI will have the unique ability to triangulate the dark energy properties and avoid pitfalls arising from systematics of any one or two methods, thereby unraveling the nature of dark energy with accuracy and precision.